\documentclass[journal]{IEEEtran}
\usepackage{amsmath}
\usepackage{graphicx}
\usepackage{caption}
\usepackage{amsfonts}
\usepackage{color}
\usepackage{subcaption}

\usepackage{makecell} 
\usepackage{multirow}
\usepackage{float}
\usepackage{xcolor}
\usepackage{comment}
\usepackage{amsmath, amssymb}
\usepackage{amsthm}
\theoremstyle{plain} 
\newtheorem{theorem}{Theorem}

\usepackage{graphicx}
\usepackage{array}
\usepackage{booktabs}
\usepackage{algorithm}
\usepackage{algorithmic}
\IEEEoverridecommandlockouts
\begin{document}

\title{Ambiguity-Optimized Mirabbasi-Martin Filter for Radar Sensing Using Zak-OTFS}
\author{Naveed Bin Nazir and Ananthanarayanan Chockalingam \\
Department of ECE, Indian Institute of Science, Bangalore 560012
}
\maketitle

\begin{abstract}
The radar sensing performance of Zak transform-based orthogonal time frequency space (Zak-OTFS) waveform is strongly influenced by its self-ambiguity function, which  depends on the delay-Doppler (DD) pulse shaping filter used. In this letter, we propose a pulse shaping filter in the Mirabbasi-Martin (MM) filter design framework and optimize the ambiguity function of the waveform in such a way that the sidelobe levels of the ambiguity function are substantially lowered while keeping the main-lobe width (MLW) narrow. In the proposed ambiguity-optimized MM (AO-MM) filter, the coefficients are tuned to jointly minimize the normalized MLW and integrated sidelobe ratio (ISLR), without incurring time or bandwidth expansion. Our results show that the proposed AO-MM filter achieves significantly better range and velocity estimation performance compared to those of the benchmark sinc, Gaussian, and Gaussian-sinc filters in densely as well as sparsely populated multi-target scenes.   
\end{abstract}

\section{Introduction}
\label{sec:intro}
Zak transform-based orthogonal time frequency space (Zak-OTFS) waveform uses a delay-Doppler (DD) domain pulse as the basic carrier, realized in time as a pulsone (tone modulated pulse train) via inverse Zak transform \cite{zak_otfs3}. Its predictable and non-fading DD domain input-output relation makes it robust to channels with large delay and Doppler spreads, and hence attractive for communication in high-mobility channels. Inherently operating in the DD domain, Zak-OTFS also extends naturally to radar sensing \cite{radar1}. Its sensing performance is governed by its self-ambiguity function, which is shaped by the DD pulse shaping filter used. The distribution of the ambiguity volume around the origin is characterized by three metrics: $i)$ the main-lobe width (MLW) at $-25$~dB level, which dictates the resolvability of closely spaced targets, $ii)$ the peak sidelobe ratio (PSLR), i.e., the ratio of the highest sidelobe peak to the main-lobe peak, and $iii)$ the integrated sidelobe ratio (ISLR), i.e., the ratio of the area under the sidelobes to that under the main lobe, which governs the masking of weak targets by strong echoes \cite{radar2}.

The sensing performance of Zak-OTFS with different pulse shaping filters such as sinc filter, Gaussian filter \cite{zak_otfs4}, and Gaussian-sinc (GS) filter \cite{zak_otfs5} has been studied in \cite{radar2}. It has been shown that, among these filters, the sinc filter has the narrowest ambiguity function main lobe, achieving the best performance in scenarios with  densely populated targets. On the other hand, due to the very small footprint of the ambiguity function with negligible sidelobes, Gaussian filter outperforms in sparsely populated scenarios. The GS filter's ambiguity function and hence its performance are similar to those of the sinc filter. This varied sensing performance of different filters in sparsely and densely populated scenes 
motivates the search for filters that can achieve uniformly better performance across both sparse and dense scenes. In this letter, we address this issue and propose an ambiguity-optimized filter in the Mirabbasi-Martin (MM) filter design framework which achieves superior sensing performance results.

The MM filter was originally designed as a prototype filter for transmultiplexers/filter banks \cite{MM_1} to achieve strong stopband attenuation in the frequency domain. Recently, in \cite{MM_zak}, the MM prototype filter has been extended to 
the DD domain for Zak-OTFS, where the coefficients of the filter were optimized for achieving improved communication performance with higher-order QAM. In this letter, we consider the MM filter design framework for radar sensing. Our approach is to directly optimize the ambiguity function that governs radar sensing performance. While the optimization for communication in \cite{MM_zak} works on the shape of the MM pulse, the optimization for radar sensing in this letter works directly on the ambiguity function of the MM pulse. We optimize the coefficients of the MM prototype filter in such a way that the sidelobe levels in the ambiguity function are substantially lowered while keeping the main-lobe width narrow, without incurring time or bandwidth expansion. The novel contributions in this letter are summarized as follows.

\begin{itemize}
\item We propose an ambiguity-optimization framework for the MM filter coefficients, which jointly minimizes the normalized MLW and ISLR in the ambiguity function under the $99\%$ energy-containment criterion. The resulting ambiguity-optimized MM (AO-MM) filter has an MLW of $3.60$ delay bins, significantly narrower than the $5.39$ bins of the Gaussian filter, and an ISLR of $-21.0$\,dB, far below the $2.31$\,dB and $-3.12$\,dB of the sinc and GS filters, respectively.
\item We derive a closed-form expression for the ambiguity function of the MM-filtered Zak-OTFS waveform, which enables exact evaluation of the design objective and accelerates the simulations by a factor of about $70$ compared to numerical integration.
\item We evaluate the sensing performance of the proposed AO-MM filter in two-target and multi-target scenes, covering the regimes of sparsely and densely populated scenarios. Our simulation results show that the AO-MM filter achieves significantly better range and velocity estimation performance compared to those of the benchmark sinc, Gaussian, and GS filters in sparsely as well as densely populated target scenarios. 
\end{itemize}

\section{Zak-OTFS Radar System Model}
\label{sec:system}
\begin{figure}[t]
\includegraphics[width=9cm, height=3.75cm]{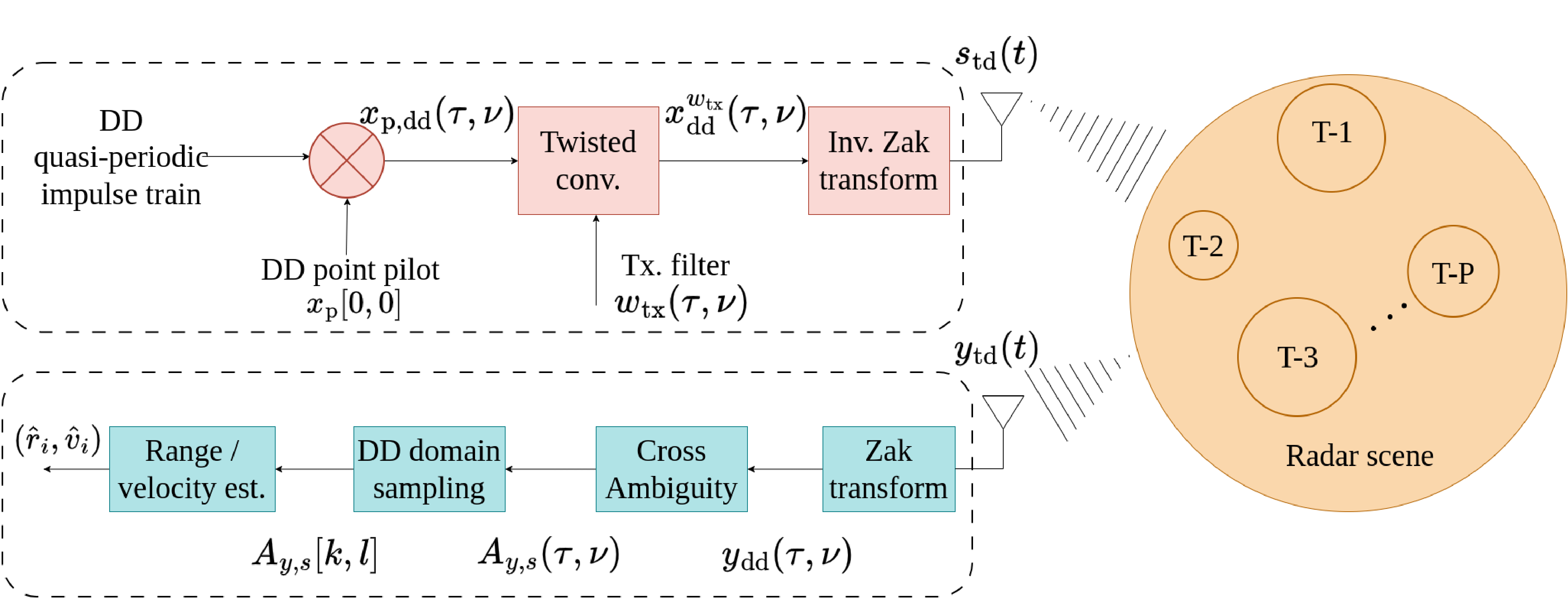}
\caption{Zak-OTFS radar sensing system.} 
\label{fig:zak_otfs_block_diagram}
\end{figure}

Fig.~\ref{fig:zak_otfs_block_diagram} shows the block diagram of a Zak-OTFS radar system. In Zak-OTFS radar, the probing waveform is a pulse in the DD domain, i.e., a quasi-periodic function with delay period $\tau_{\text{p}}$ and Doppler period $\nu_{\text{p}}=1/\tau_{\text{p}}$. The fundamental region $\mathcal{D}_{0}=[0,\tau_{\text{p}})\times[0,\nu_{\text{p}})$ is divided into $M=B\tau_{\text{p}}$ delay bins and $N=T\nu_{\text{p}}$ Doppler bins, where $B$ and $T$ are the bandwidth and duration of the waveform. A unit-energy DD impulse $x_{\text{p}}[k,l]=\delta[k,l]$, $0\leq k<M$, $0\leq l<N$, is made quasi-periodic and converted to a continuous DD signal as
\begin{eqnarray}
\label{eq:xpdd}
x_{\text{p,dd}}(\tau,\nu)=\sum_{k,l\in\mathbb{Z}}x_{\text{p,dd}}[k,l]\,\delta\Big(\tau-\frac{k\tau_{\text{p}}}{M}\Big)\delta\Big(\nu-\frac{l\nu_{\text{p}}}{N}\Big),
\end{eqnarray}
where $x_{\text{p,dd}}[k+nM,l+mN]=x_{\text{p}}[k,l]e^{j2\pi nl/N}$, $n,m\in\mathbb{Z}$. To limit the time and bandwidth, a separable DD pulse shaping filter $w_{\text{tx}}(\tau,\nu)=w_{\tau}(\tau)w_{\nu}(\nu)$ is applied, giving $x^{w_{\text{tx}}}_{\text{dd}}(\tau,\nu)=w_{\text{tx}}(\tau,\nu)*_{\sigma}x_{\text{p,dd}}(\tau,\nu)$. The transmitted waveform is $s_{\mathrm{td}}(t)=Z_{t}^{-1}(x^{w_{\text{tx}}}_{\text{dd}}(\tau,\nu))$, where $*_{\sigma}$ and $Z_{t}^{-1}$ denote twisted convolution and inverse Zak transform, respectively\footnote{$a(\tau,\nu)*_{\sigma}b(\tau,\nu)=\int\int a(\tau',\nu')b(\tau-\tau',\nu-\nu')e^{j2\pi\nu'(\tau-\tau')}d\tau'd\nu'$, $Z_{t}(a(t))=\sqrt{\tau_{\text{p}}}\sum_{k\in\mathbb{Z}}a(\tau+k\tau_{\text{p}})e^{-j2\pi\nu k\tau_{\text{p}}}$, $Z_{t}^{-1}(a(\tau,\nu))=\sqrt{\tau_{\text{p}}}\int_{0}^{\nu_{\text{p}}}a(t,\nu)d\nu$.}. The filter is normalized such that $s_{\mathrm{td}}(t)$ has unit energy.

The sensing scene is characterized by the DD spreading function $h_{\text{s}}(\tau,\nu)=\sum_{i=1}^{P}h_{i}\delta(\tau-\tau_{i})\delta(\nu-\nu_{i})$, where $P$ is the number of targets, and $\tau_{i}$, $\nu_{i}$ and $h_{i}$ are the delay, Doppler and complex gain of $i$th target. The received echo signal is 
\begin{equation}
\label{eq:echo}
y_{\mathrm{td}}(t)=\sum\nolimits_{i=1}^{P}h_{i}\,s_{\mathrm{td}}(t-\tau_{i})\,e^{j2\pi\nu_{i}(t-\tau_{i})}+n_{\mathrm{td}}(t),
\end{equation}
where $n_{\mathrm{td}}(t)$ is additive white Gaussian noise (AWGN) with $\mathbb{E}[n_{\mathrm{td}}(t)n_{\mathrm{td}}^{*}(t-t')]=N_{0}\delta(t')$. The receiver converts the echo to the DD domain as $y_{\text{dd}}(\tau,\nu)=Z_{t}(y_{\mathrm{td}}(t))$ and computes its cross-ambiguity with the transmitted waveform \cite{zak_otfs3}:
\begin{eqnarray}
\label{eq:Ayx}
A_{y,s}(\tau,\nu) & \hspace{-2mm} = & \hspace{-2mm} \int y_{\mathrm{td}}(t)s_{\mathrm{td}}^{*}(t-\tau)e^{-j2\pi\nu(t-\tau)}dt \nonumber \\
& \hspace{-34mm} = & \hspace{-19mm} \int\hspace{-1mm}\int_{\mathcal{D}_{0}} y_{\text{dd}}(\tau',\nu')\,x^{w_{\text{tx}}*}_{\text{dd}}(\tau'-\tau,\nu'-\nu)\,e^{-j2\pi\nu(\tau'-\tau)}d\tau'd\nu' \nonumber \\
& \hspace{-34mm} = & \hspace{-19mm} \sum\nolimits_{i=1}^{P}h_{i}A_{s,s}(\tau-\tau_{i},\nu-\nu_{i})e^{j2\pi\nu_{i}(\tau-\tau_{i})}+A_{n,s}(\tau,\nu),
\end{eqnarray}
where $A_{s,s}(\tau,\nu)$ and $A_{n,s}(\tau,\nu)$ follow the same definition with $y_{\mathrm{td}}(t)$ replaced by $s_{\mathrm{td}}(t)$ and $n_{\mathrm{td}}(t)$, respectively. Each target therefore appears in $A_{y,s}(\tau,\nu)$ as a copy of $A_{s,s}(\tau,\nu)$ centered at its DD location. The noise term satisfies $\mathbb{E}[A_{n,s}(\tau_{1},\nu_{1})A^{*}_{n,s}(\tau_{2},\nu_{2})]=N_{0}e^{-j2\pi\nu_{2}(\tau_{2}-\tau_{1})}A_{s,s}(\tau_{1}-\tau_{2},\nu_{1}-\nu_{2})$. For real and even $w_{\tau}(\tau)$ and $w_{\nu}(\nu)$, which holds for the filters considered in this letter, the self-ambiguity of $s_{\mathrm{td}}(t)$ is given by \cite{radar2}
\begin{eqnarray}
\label{eq:Axx}
A_{s,s}(\tau,\nu) & \hspace{-2mm} = & \hspace{-2mm} \sum_{n,m\in\mathbb{Z}}\int w_{\tau}(\tau')w_{\tau}(t_{n}-\tau')e^{j2\pi\nu(t_{n}-\tau')}d\tau' \nonumber \\
& & \hspace{4mm} \times\int w_{\nu}(\nu')w_{\nu}(f_{m}-\nu')e^{j2\pi n\tau_{\text{p}}\nu'}d\nu',
\end{eqnarray}
where $t_{n}=\tau-n\tau_{\text{p}}$ and $f_{m}=\nu-m\nu_{\text{p}}$. The delay and Doppler resolutions are $1/B$ and $1/T$, corresponding to range and velocity resolutions of $c/(2B)$ and $c/(2f_{c}T)$, with speed of light $c$ and carrier frequency $f_{c}$ \cite{radar1}.

{\em MM pulse shaping filter:} In this letter, we consider MM pulse shaping filter and optimize its coefficients for radar sensing. The MM filter in the DD domain is defined by its delay and Doppler pulses, which, for $\xi\in\{\tau,\nu\}$ with $X_{\tau}=B$ and $X_{\nu}=T$, are given by
\begin{eqnarray}
\label{eq:mm_pulse}
w_{\xi}(x) & \hspace{-2mm} = & \hspace{-2mm} \kappa_{\xi}\sqrt{X_{\xi}}\,p_{\xi}(x)\,\text{rect}\Big(\frac{X_{\xi}x}{\zeta_{\xi}}\Big), \nonumber \\
p_{\xi}(x) & \hspace{-2mm} = & \hspace{-2mm} c_{0}+2\sum\nolimits_{\ell=1}^{g-1}c_{\ell}\cos\Big[\frac{2\pi\ell}{\zeta_{\xi}}\Big(X_{\xi}x+\frac{\zeta_{\xi}}{2}\Big)\Big].
\end{eqnarray}
The real, even pulse $w_\xi(x)$ is shaped by the $g$ MM coefficients $\{c_{0},c_{1},\ldots,c_{g-1}\}$, where $g$ is the overlap factor. Here, $\kappa_{\xi}$ normalizes $w_{\xi}(x)$ to unit energy, and the roll-off factors $\zeta_{\tau}$ and $\zeta_{\nu}$ ensure that $99\%$ of the energy lies within $B$ and $T$. While these coefficients were optimized for communication in \cite{MM_zak}, we optimize them for radar sensing in Sec. \ref{sec:proposed}.

{\em Crystallization condition:} $A_{s,s}(\tau,\nu)$ has replicas centered at every point of the period lattice $\Lambda_{\mathrm{p}}=\{(n\tau_{\mathrm{p}},m\nu_{\mathrm{p}})\,|\,n,m\in\mathbb{Z}\}$. Only the central response at $(0,0)$ is of interest, while the remaining replicas are DD aliases that may appear as ghost targets. Let $\tau_{\text{min}}$ and $\tau_{\text{max}}$ ($\nu_{\text{min}}$ and $\nu_{\text{max}}$) denote the minimum and maximum target delays (Dopplers). The aliases do not overlap the true target peaks when $\tau_{\text{p}}>(\tau_{\text{max}}-\tau_{\text{min}})$ and $\nu_{\text{p}}>(\nu_{\text{max}}-\nu_{\text{min}})$, termed the crystallization condition \cite{radar1}.

{\em Peak detection:} The receiver evaluates $A_{y,s}(\tau,\nu)$ on a DD grid oversampled by factors $Q_{\tau}$ and $Q_{\nu}$ along delay and Doppler, i.e., $A_{y,s}[k,l]=A_{y,s}(\frac{k}{Q_{\tau}B},\frac{l}{Q_{\nu}T})$. The evaluation is restricted to a detection window, which encloses all targets with a margin for the spread of $A_{s,s}(\tau,\nu)$ and spans less than $\tau_{\text{p}}$ and $\nu_{\text{p}}$. Within this window, the locations of the $P$ largest local maxima of $|A_{y,s}[k,l]|$ are taken as the detected target peaks. The range and velocity estimates of the $i$th target are obtained from its detected peak $(\hat{k}_{i},\hat{l}_{i})$ as
\begin{eqnarray}
\label{eq:est}
\hat{r}_{i}=\frac{c\,\hat{k}_{i}}{2Q_{\tau}B}, \ \ \ \hat{v}_{i}=\frac{c\,\hat{l}_{i}}{2f_{c}Q_{\nu}T}.
\end{eqnarray}

\section{Proposed Ambiguity-Optimized MM filter}
\label{sec:proposed}

\subsection{Ambiguity-Optimization of MM Coefficients}
\label{sub:opt}

A narrow main lobe (sinc filter) resolves closely spaced targets, while a low ISLR (Gaussian filter) prevents strong echoes from masking weak ones. The MM coefficients offer direct control over this trade-off. Unlike \cite{MM_zak}, which optimizes them for DD-pulse localization, we optimize them directly on the self-ambiguity function.

\begin{figure}[t]
\centering
\begin{subfigure}[b]{0.49\linewidth}
    \includegraphics[width=\linewidth,height=3.6cm]{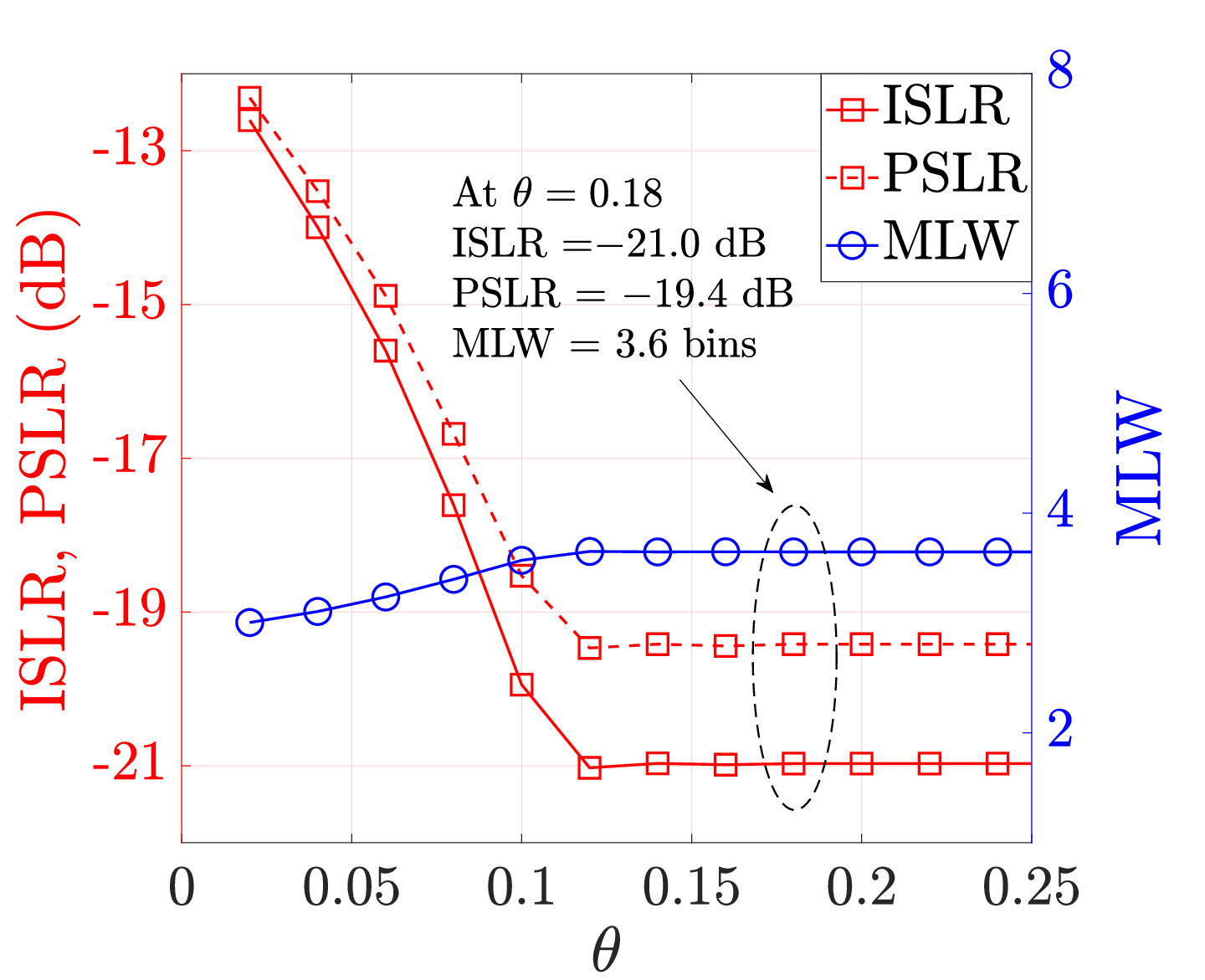}
    \label{fig:2a}
    \vspace{-4mm}
    \caption{}
\end{subfigure}
\begin{subfigure}[b]{0.49\linewidth}
    \includegraphics[width=\linewidth,height=3.6cm]{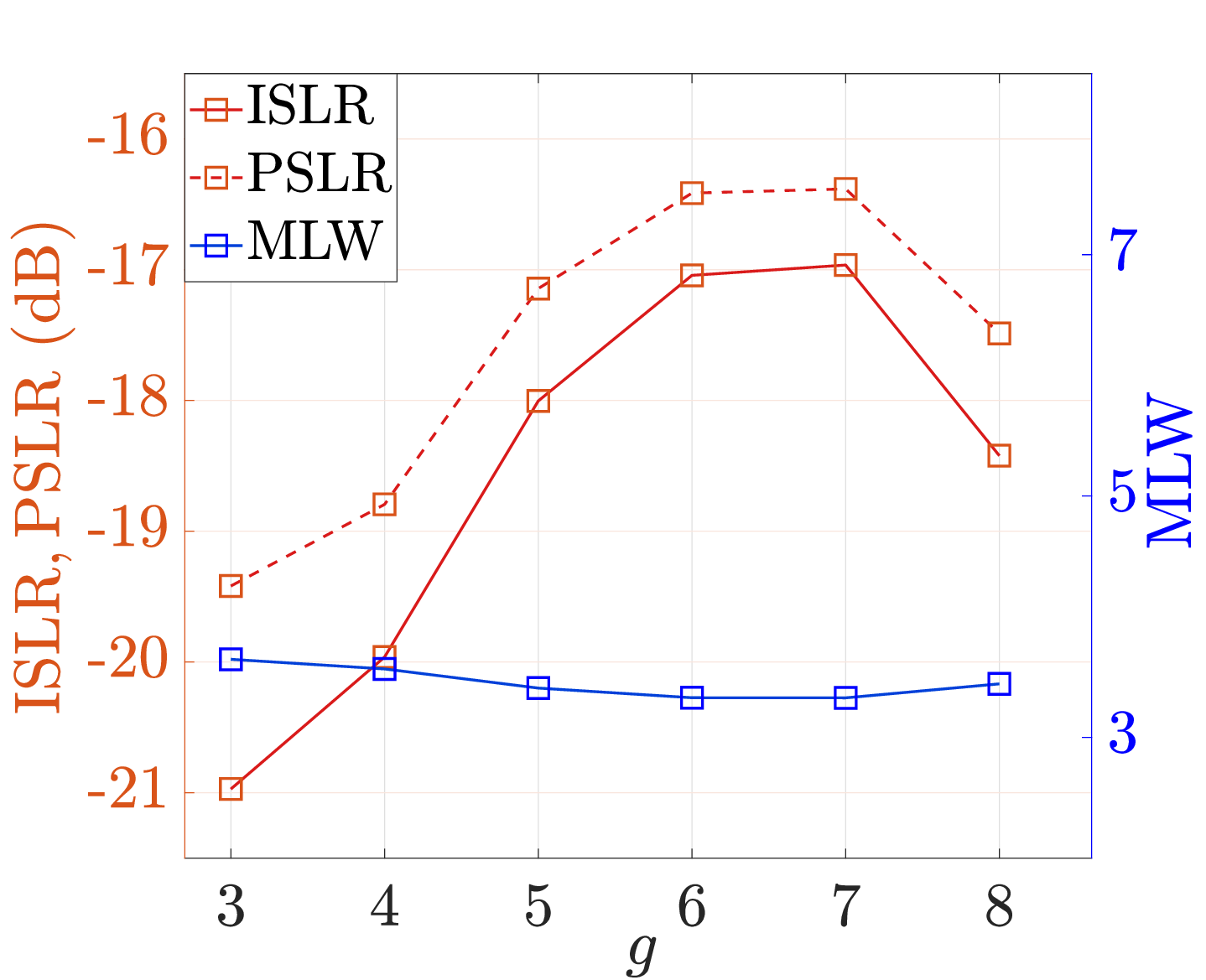}
    \label{fig:2b}
    \vspace{-4mm}
    \caption{}
\end{subfigure}
\caption{ISLR, PSLR, and MLW of the AO-MM filter: (a) as a function of $\theta$ for $g=3$; (b) as a function of $g$, each at its selected $\theta$.}
\label{Fig2}
\end{figure}

For a given $g$, let $\mathbf{c}^{(g)}_{\mathrm{MM}}=[c_0,c_1,\ldots,c_{g-1}]^T$ denote the original prototype MM coefficient vector of \cite{MM_1}, and consider the modified coefficient vector
\begin{equation}
\mathbf{c}^{(g)}=\mathbf{c}^{(g)}_{\text{MM}}+\boldsymbol{\eta}^{(g)},
\label{eq:c_pert}
\end{equation}
where $\boldsymbol{\eta}^{(g)}=[0,\eta_{1},\ldots,\eta_{g-1}]^{T}$ keeps $c^{(g)}_{0}=1$ fixed \cite{MM_zak}. For each candidate $\mathbf{c}^{(g)}$, the roll-off factors $\zeta_{\tau}$ and $\zeta_{\nu}$ are recomputed to satisfy the $99\%$ energy-containment criterion within $B$ and $T$, while $\kappa_{\tau}$ and $\kappa_{\nu}$ are adjusted to ensure unit-energy. Thus, no candidate gains resolution through time or bandwidth expansion. The design metrics are evaluated on the zero-Doppler cut\footnote{Since the delay and Doppler pulses use the same coefficients and energy criterion, they have identical shapes in the normalized variables $B\tau$ and $T\nu$. Hence, the zero-delay cut $|A_{s,s}(0,\nu)|$ versus $T\nu$ coincides with the zero-Doppler cut $|A_{s,s}(\tau,0)|$ versus $B\tau$, and optimizing one optimizes both.} $|A_{s,s}(\tau,0)|$ of the central response. Let $\mathrm{MLW}(\mathbf{c}^{(g)})$, measured in delay bins of $1/B$, and $\mathrm{ISLR}(\mathbf{c}^{(g)})$, expressed on a linear scale, denote the corresponding metrics. We define the objective
\begin{equation}
J\big(\mathbf{c}^{(g)}\big)\triangleq
\frac{\mathrm{MLW}\big(\mathbf{c}^{(g)}\big)}{\mathrm{MLW}\big(\mathbf{c}^{(g)}_{\text{MM}}\big)}
+\frac{\mathrm{ISLR}\big(\mathbf{c}^{(g)}\big)}{\mathrm{ISLR}\big(\mathbf{c}^{(g)}_{\text{MM}}\big)},
\label{eq:J}
\end{equation}
and formulate the coefficient optimization problem as
\begin{equation}
\begin{aligned}
\underset{\mathbf{c}^{(g)}}{\text{minimize}}\quad & J\big(\mathbf{c}^{(g)}\big)\\
\text{subject to}\quad & c^{(g)}_{0}=1,\\
& \big|c^{(g)}_{i}-c^{(g)}_{\text{MM},i}\big|\leq\theta,\ \ i=1,\ldots,g-1,
\end{aligned}
\label{eq:opt}
\end{equation}
where $\theta$ bounds the admissible coefficient perturbations. The normalization in (\ref{eq:J}) makes both terms dimensionless and equal to unity at $\mathbf{c}^{(g)}_{\text{MM}}$, trading relative main lobe expansion one-to-one against relative sidelobe reduction. The non-smooth PSLR is omitted from (\ref{eq:J}) as it naturally follows the ISLR (see Fig.~\ref{Fig2}). 

Problem (\ref{eq:opt}) is non-convex and is solved via sequential quadratic programming (SQP) for $\theta$ on a grid with step $0.02$. At each $\theta$, the solver is initialized at both $\boldsymbol{\eta}^{(g)}=\mathbf{0}$ and the previous $\theta$, and the better result is retained. To keep the SQP gradients free of any numerical-integration errors, the zero-Doppler cut is evaluated exactly as $A_{s,s}(\tau,0)$ for $|\tau|\le\tau_{\mathrm{p}}/2$, using the closed-form expression derived in Sec.~\ref{sub:closed}. 

\begin{table}[t]
\centering
\begin{tabular}{|c||c|c|}
\hline
Parameter & MM (unopt.) & AO-MM (prop.) \\ \hline\hline
$\mathbf{c}^{(3)}$ & $[1,-0.9114,0.4114]^{T}$ & $[1,-0.8041,0.3471]^{T}$ \\ \hline
$\zeta_{\tau}=\zeta_{\nu}$ & $4.3913$ & $4.3671$ \\ \hline
$\kappa_{\tau}=\kappa_{\nu}$ & $0.2755$ & $0.3006$ \\ \hline
\end{tabular}
\caption{Parameters of the unoptimized MM filter and the proposed AO-MM filter.}
\label{tab:opt_params}
\end{table}
\begin{figure}[t]
\centering
\begin{subfigure}[b]{0.49\linewidth}
    \includegraphics[width=\linewidth,height=3.6cm]{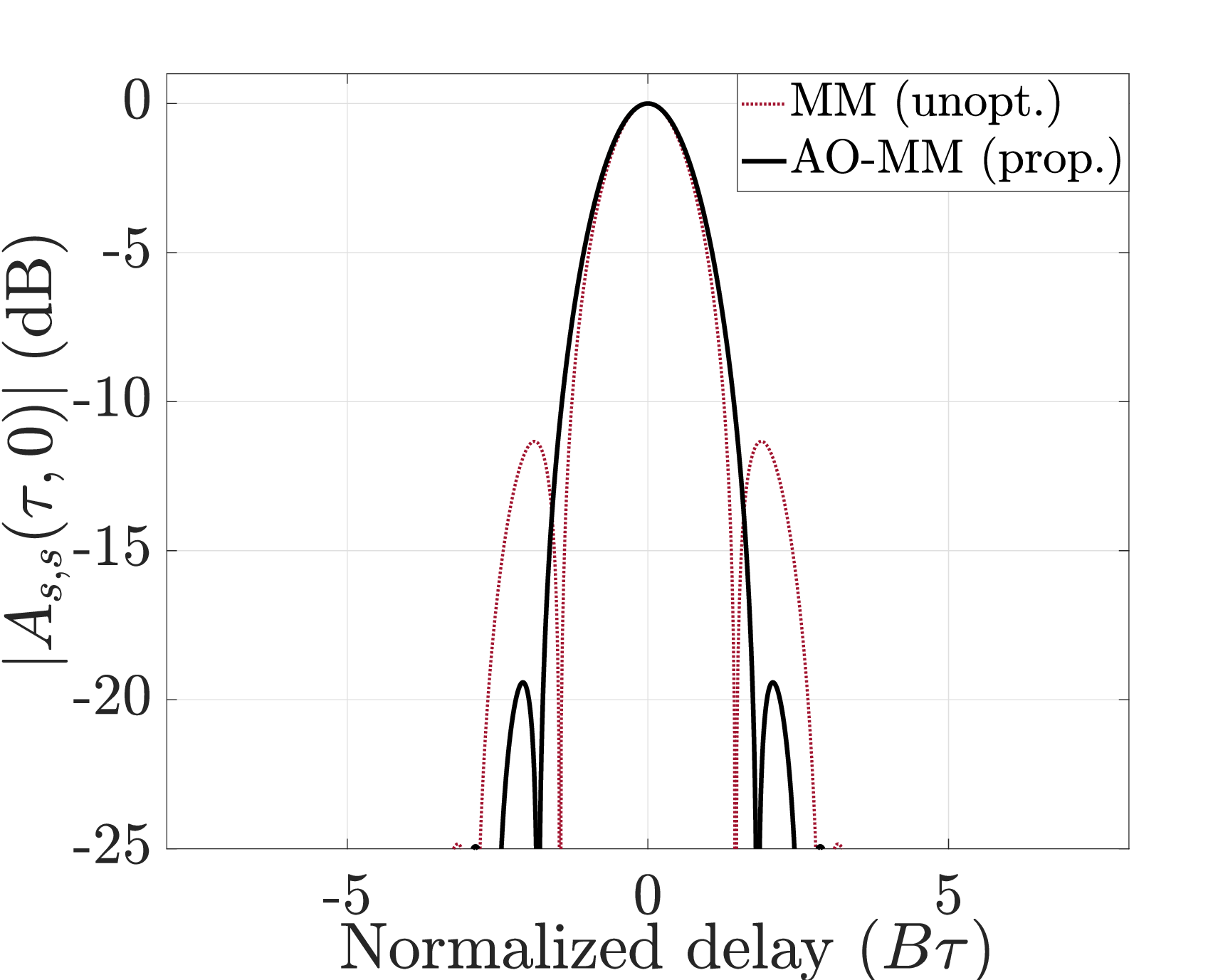}
    \label{Fig:3a}
    \vspace{-4mm}
    \caption{}
\end{subfigure}
\begin{subfigure}[b]{0.49\linewidth}
    \includegraphics[width=\linewidth,height=3.6cm]{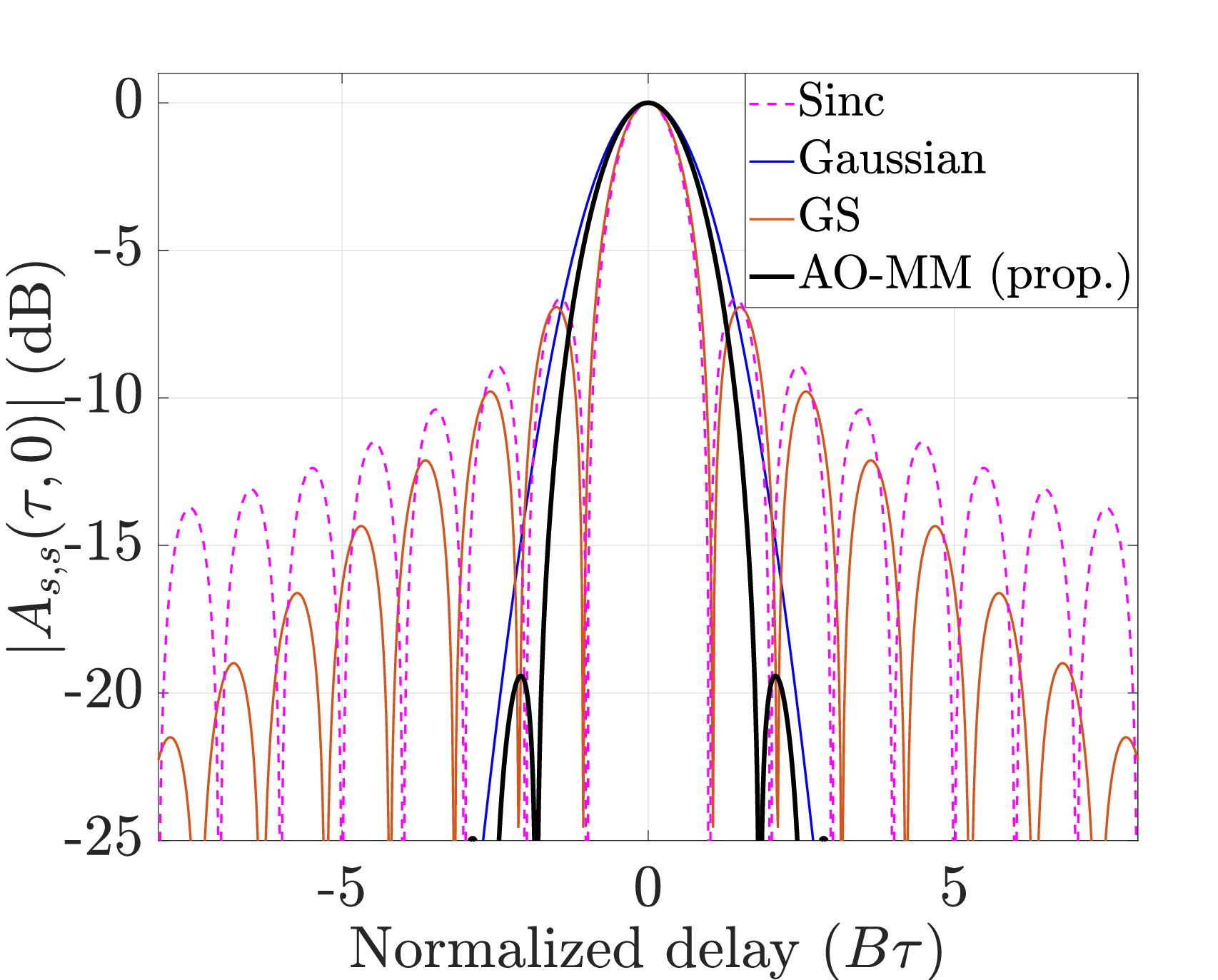}
    \label{Fig:3b}
    \vspace{-4mm}
    \caption{}
\end{subfigure}
\caption{Zero-Doppler cut $|A_{s,s}(\tau,0)|$: (a) MM (unopt.) and AO-MM (prop.); (b) sinc, Gaussian, GS, and AO-MM (prop.).}
\label{Fig3}
\end{figure}
\begin{table}[t]
\centering
\begin{tabular}{|c||c|c|c|c|c|}
\hline
Metric & Sinc & Gaussian & GS & MM (unopt.) & AO-MM (prop.) \\ \hline\hline
MLW (bins) & $1.99$ & $5.39$ & $2.12$ & $2.92$ & $3.60$ \\ \hline
PSLR (dB) & $-6.65$ & $-$ & $-6.92$ & $-11.40$ & $-19.40$ \\ \hline
ISLR (dB) & $2.31$ & $-$ & $-3.12$ & $-11.30$ & $-21.00$ \\ \hline
\end{tabular}
\caption{Self-ambiguity characteristics of different filters.}
\label{tab:af_metrics}
\end{table}
{\em Selection of $\theta$ and $g$:} The bound $\theta$ regulates coefficient deviation from the MM prototype. Fig.~\ref{Fig2}(a) shows that increasing $\theta$ reduces ISLR at the expense of a wider main lobe, revealing a leakage-localization trade-off. The metrics remain essentially unchanged for $\theta\geq0.12$, and $\theta=0.18$ with the lowest $J$ is chosen, yielding an ISLR of $-21.0$\,dB, a PSLR of $-19.4$\,dB and an MLW of $3.6$ delay bins. For other values of $g$, $\theta$ is selected in the same manner. Fig.~\ref{Fig2}(b) shows that MLW remains nearly constant across $g$, while $g=3$ achieves the lowest ISLR and PSLR, making it the adopted choice for the proposed filter.
For $g=3$ and $\theta=0.18$, (\ref{eq:opt}) yields
\begin{equation}
\mathbf{c}^{(3)}_{\text{AO}}=[1,\,-0.8041,\,0.3471]^{T}.
\label{eq:c_AO}
\end{equation}
Substituting $\mathbf{c}^{(3)}_{\text{AO}}$ into (\ref{eq:mm_pulse}) defines the proposed AO-MM filter, whose parameters are compared against the unoptimized MM filter in Table~\ref{tab:opt_params}.

\subsection{Ambiguity Characteristics of the AO-MM Filter}
\label{sub:char}
Fig.~\ref{Fig3}(a) compares the zero-Doppler cuts of the self-ambiguity functions for the MM (unopt.) and AO-MM filters ($g=3$). The proposed optimization substantially suppresses sidelobe leakage at the expense of a slightly wider main lobe, which is the balance that minimizes (\ref{eq:J}). Fig.~\ref{Fig3}(b) compares the AO-MM filter against the benchmark sinc, Gaussian, and GS filters, with the corresponding metrics summarized in Table~\ref{tab:af_metrics}\footnote{Following \cite{radar2}, $|A_{s,s}(\tau,0)|$ is normalized to its peak, levels in dB are $10\log_{10}|A_{s,s}|$ and the ISLR is evaluated over $\tau\in[-\tau_p/2,\tau_p/2]$ throughout.}. The main lobe of the AO-MM filter is about $33\%$ narrower than that of the Gaussian filter, while its PSLR and ISLR are more than $12$\,dB and $17$\,dB below those of the sinc and GS filters. Moreover, owing to the compact DD support of the MM pulse, $A_{s,s}(\tau,0)$ vanishes for $|B\tau|\ge\zeta_\tau\approx4.37$, so no sidelobe energy leaks beyond a few delay bins. 

\subsection{Closed-Form Self-Ambiguity Function}
\label{sub:closed}
Evaluating $A_{s,s}(\tau,\nu)$ in (\ref{eq:Axx}) via numerical integration on a fine DD grid is computationally costly. We therefore derive a closed-form expression for $A_{s,s}(\tau,\nu)$. For $\xi\in\{\tau,\nu\}$, define $\rho_{\xi}\triangleq X_{\xi}/\zeta_{\xi}$ and the overlap length $\Delta_{\xi}(u)\triangleq\big[1/\rho_{\xi}-|u|\big]^{+}$, where $[x]^{+}\triangleq\max(x,0)$.  
Define the finite-series coefficients
\begin{equation}
a_{z}=
\begin{cases}
c_{0}, & z=0,\\
(-1)^{|z|}c_{|z|}, & 1\le |z|\le g-1.
\end{cases}
\label{eq:a_coeff}
\end{equation}

\begin{theorem}
\label{thm:AF}
The self-ambiguity function of the Zak-OTFS waveform with the MM filter in (\ref{eq:mm_pulse}) is given by
\begin{equation}
A_{s,s}(\tau,\nu)
=\sum_{n\in\mathbb{Z}}\sum_{m\in\mathbb{Z}}
\Psi_{\tau}\big(\tau-n\tau_{\mathrm{p}},\,\nu\big)\,
\Psi_{\nu}\big(\nu-m\nu_{\mathrm{p}},\,n\tau_{\mathrm{p}}\big),
\label{eq:AF_thm}
\end{equation}
where, for $\xi\in\{\tau,\nu\}$,
\begin{eqnarray}
\hspace{-7mm}
\Psi_{\xi}(u,f)
& \hspace{-2mm} = & \hspace{-2mm} \kappa_{\xi}^{2}\zeta_{\xi}\rho_{\xi}\,\Delta_{\xi}(u)
\sum_{k,q=-(g-1)}^{g-1}a_{k}a_{q}
\nonumber\\
& \hspace{-10mm} & \hspace{-10mm} \times
e^{j\pi\left(\rho_{\xi}(k+q)+f\right)u}\,
\operatorname{sinc}\!\big(\left(\rho_{\xi}(k-q)-f\right)\Delta_{\xi}(u)\big).
\label{eq:Psi}
\end{eqnarray}
\end{theorem}

\begin{IEEEproof}
See the Appendix.
\end{IEEEproof}

Although (\ref{eq:AF_thm}) is formally expressed over $n,m\in\mathbb{Z}$, only the lattice replicas nearest to $(\tau,\nu)$ contribute, since $\Psi_{\tau}(u,f)$ and $\Psi_{\nu}(u,f)$ vanish for $|u|\geq 1/\rho_{\tau}$ and $|u|\geq 1/\rho_{\nu}$, respectively. More importantly, each contributing term is a finite sum of elementary functions rather than an integral, so $A_{s,s}(\tau,\nu)$ is evaluated 
considerably faster than by numerical integration\footnote{For $1000$ realizations on a $161\times81$ DD grid, (\ref{eq:AF_thm}) requires $0.8$\,s, versus $55.7$\,s for numerical integration (i.e., about $70\times$ faster). For the simulations, we have used MATLAB R2024b, Intel\textsuperscript{\textregistered} Core\textsuperscript{\texttrademark} Ultra~9 285K, 192\,GB RAM.}. 

\section{Results and Discussions}
\label{sec:results}
In this section, we evaluate the sensing performance of the proposed AO-MM filter and compare it with the benchmark sinc, Gaussian, and GS filters. Unless stated otherwise, all experiments use $B=4$\,MHz, $T=20$\,ms, 
$\tau_{\mathrm{p}}=100\,\mu$s, $\nu_{\mathrm{p}}=1/\tau_{\mathrm{p}}=10$\,kHz, $(M,N)=(400,200)$, $f_c=1$\,GHz, and $Q_\tau=Q_\nu=4$. Thus, one delay bin ($1/B$) corresponds to a range resolution of $c/(2B)=37.5$\,m, and one Doppler bin ($1/T$) corresponds to a velocity resolution of $c/(2f_cT)=7.5$\,m/s. The targets lie beyond $\tau_{\min}=200\,\mu$s (a range of $30$\,km), with delays $\tau_i$ measured relative to $\tau_{\min}$\footnote{Since $\tau_{\min}=2\tau_{\mathrm{p}}$, the lattice replica of the response of a target at absolute delay $\bar{\tau}_i=\tau_{\min}+\tau_i$ appears at $\tau_i$. Hence, the ranges in (\ref{eq:est}) are measured relative to $c\tau_{\min}/2$, with $\tau_{\min}$ assumed known.}, so that $\bar{\tau}_i=\tau_{\min}+\tau_i$ is the absolute delay of the $i$th target. The fades are independent with $h_i\sim\mathcal{CN}\big(0,(10^{-7}/\bar{\tau}_i)^4\big)$, and the SNR is defined for a target at $\tau_{\min}$ as $\mathrm{SNR}=(10^{-7}/\tau_{\min})^4E_s/(BTN_0)$, where $E_s=1$ is the energy of $s_{\mathrm{td}}(t)$. 

\begin{figure}[t]
\centering
\includegraphics[width=0.85\linewidth,height=0.75\linewidth]{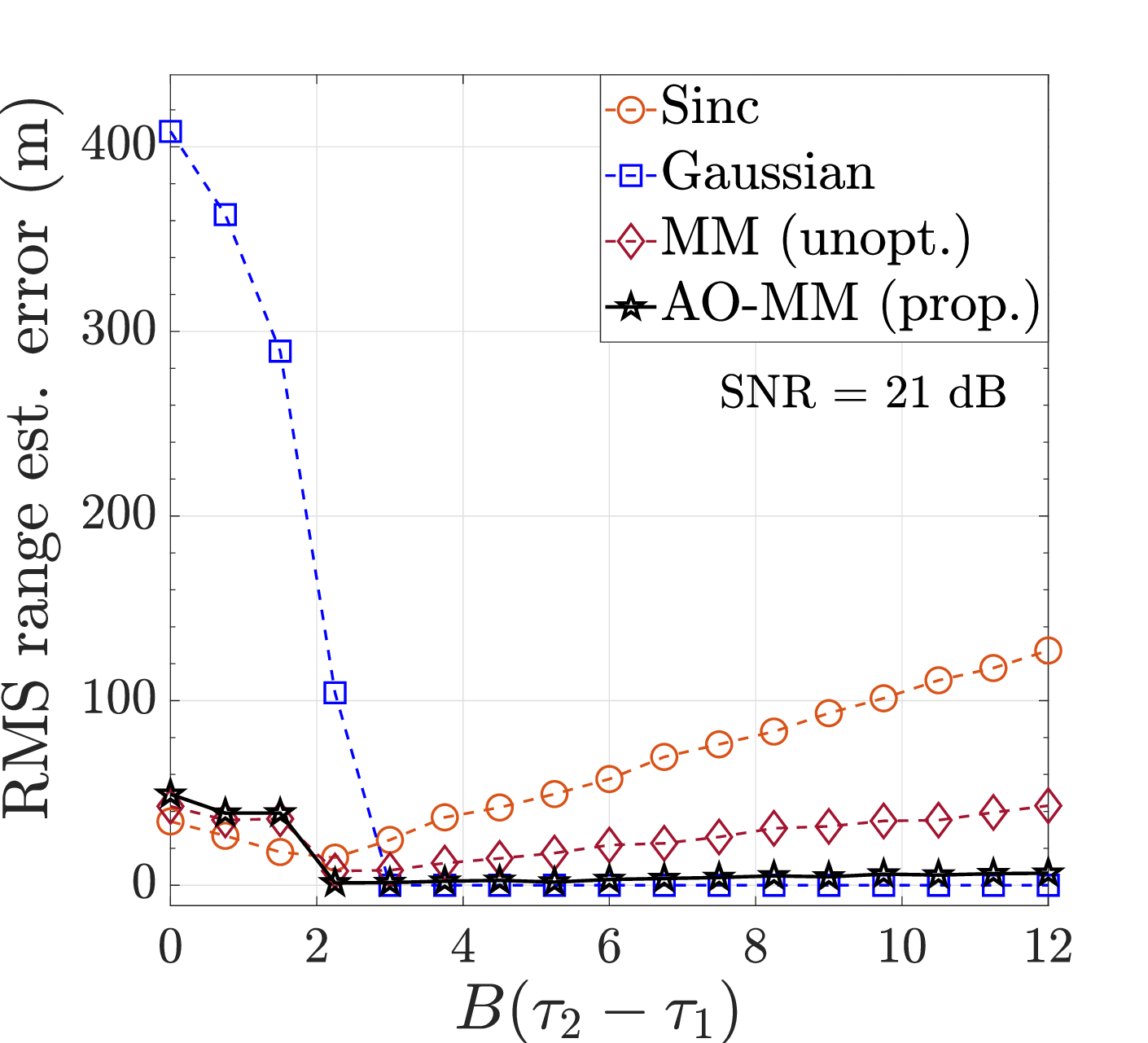}
\caption{RMS range estimation error versus normalized delay separation $B(\tau_{2}-\tau_{1})$ of two targets.}
\label{Fig4}
\end{figure}

\begin{figure}[t]
\centering
\begin{subfigure}[b]{0.49\linewidth}
    \includegraphics[width=\linewidth,height=3.6cm]{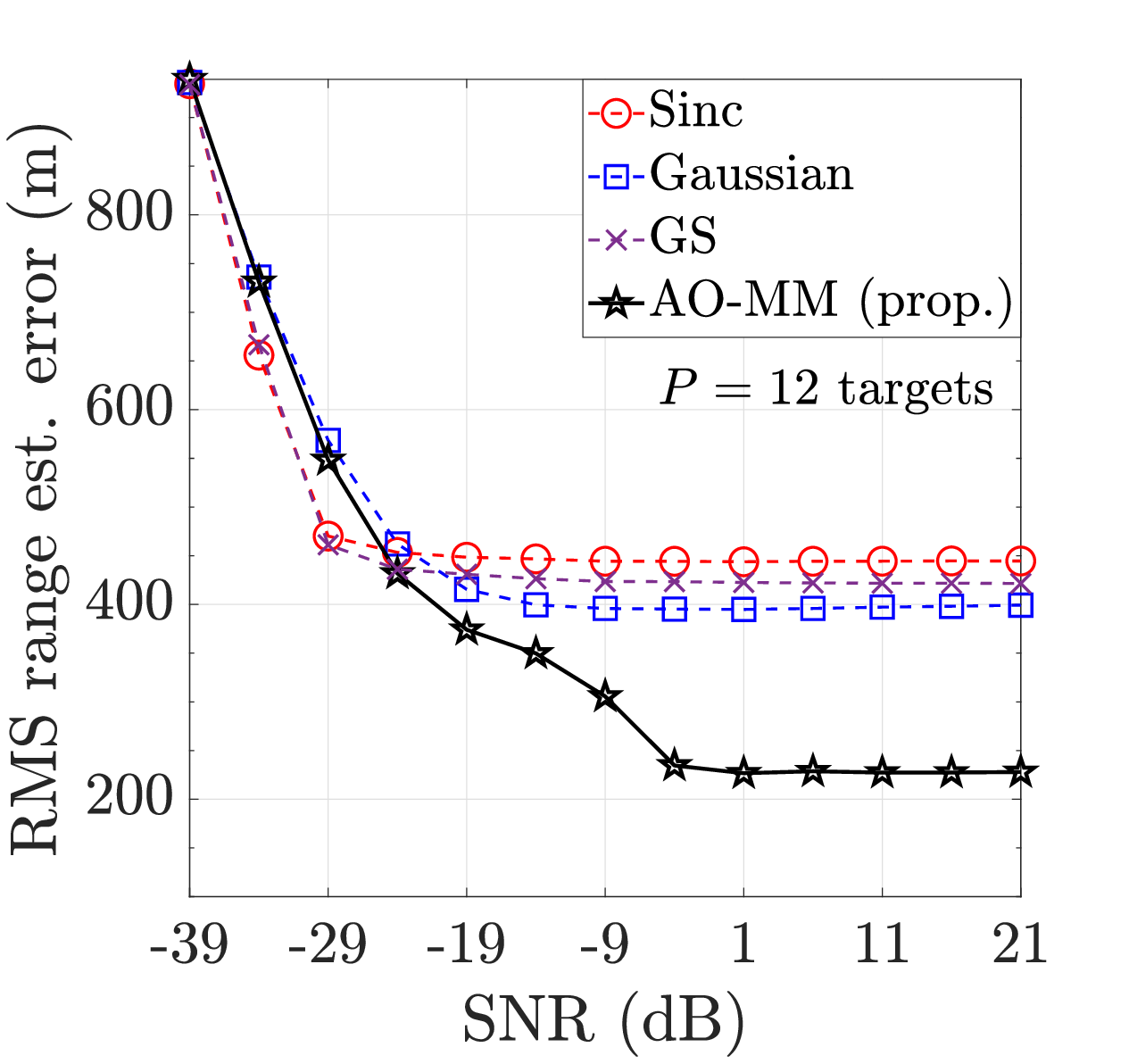}
    \label{Fig:5a}
    \vspace{-4mm}
    \caption{Range.}
\end{subfigure}
\begin{subfigure}[b]{0.49\linewidth}
    \includegraphics[width=\linewidth,height=3.6cm]{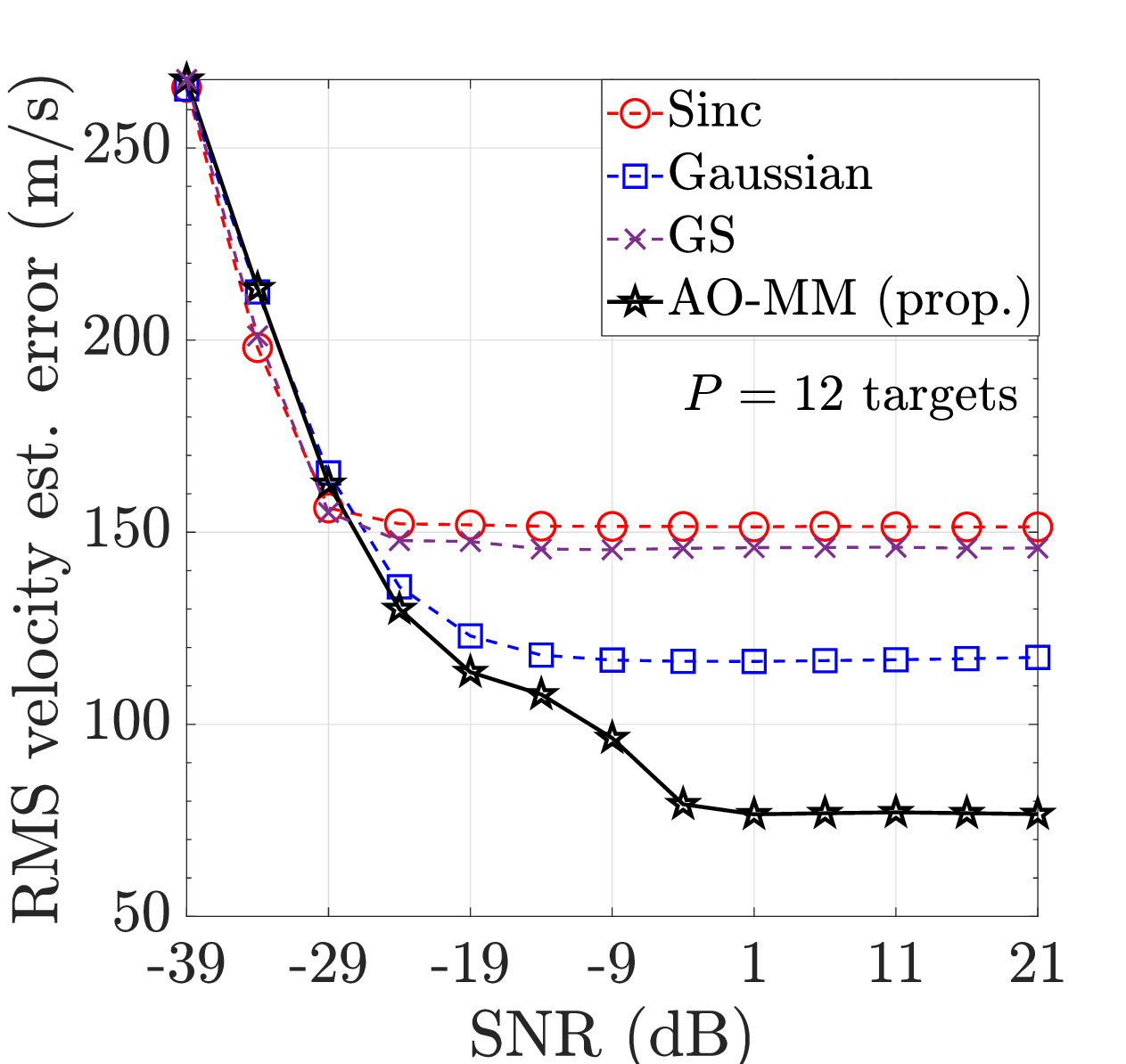}
    \label{Fig:5b}
    \vspace{-4mm}
    \caption{Velocity.}
\end{subfigure}
\caption{RMS range and velocity estimation errors versus SNR in multi-target scene with $P=12$ targets.}
\label{Fig5}
\end{figure}

{\em Two-target resolvability:} To contrast the performance of the filters for closely-spaced and well-separated targets, we consider two targets with $\nu_1=\nu_2=0$\,Hz at a high SNR of $21$\,dB. The first target is fixed at $\tau_1=0.25\,\mu$s, while $\tau_2$ is increased. The detection window is $[0,5]\,\mu$s$\,\times\,[-1,1]$\,kHz. Fig.~\ref{Fig4} shows the RMS range estimation error versus normalized delay separation $B(\tau_{2}-\tau_{1})$. Due to its narrow main lobe, sinc filter is able to resolve the targets when they are close, resulting in low estimation errors. Whereas, due to its wider main lobe, Gaussian filter tends to see the overlapped peaks as a single peak  when the targets are close, resulting in large estimation errors (e.g., when $B(\tau_2-\tau_1)<3$). 
As the separation increases, the trend reverses. The weaker echo of the far-off target can be masked by the sidelobes of the stronger echo of the nearby target. Therefore, due to its high sidelobes, sinc filter witnesses the estimation error to increase with increasing separation. Whereas, because of its almost no sidelobes, Gaussian filter finds the estimation error to drastically reduce as the separation increases. 
The unoptimized MM filter follows the behavioral pattern of the sinc filter, but its estimation error for large separation is much smaller compared to that of sinc filter, mainly due to low sidelobes compared to sinc filter.  

Importantly, the proposed AO-MM filter is observed to perform quite well with low estimation errors in both close-by and far-off target regimes. Its main lobe, only $0.68$ bins wider than that of the unoptimized MM filter, keeps its error close to that of the sinc filter for closely spaced targets, while its low PSLR and ISLR suppress masking, so that its error nearly matches the very small errors of the Gaussian filter for well-separated targets. 

\begin{figure}[t]
\centering
\begin{subfigure}[b]{0.49\linewidth}
    \includegraphics[width=\linewidth,height=3.6cm]{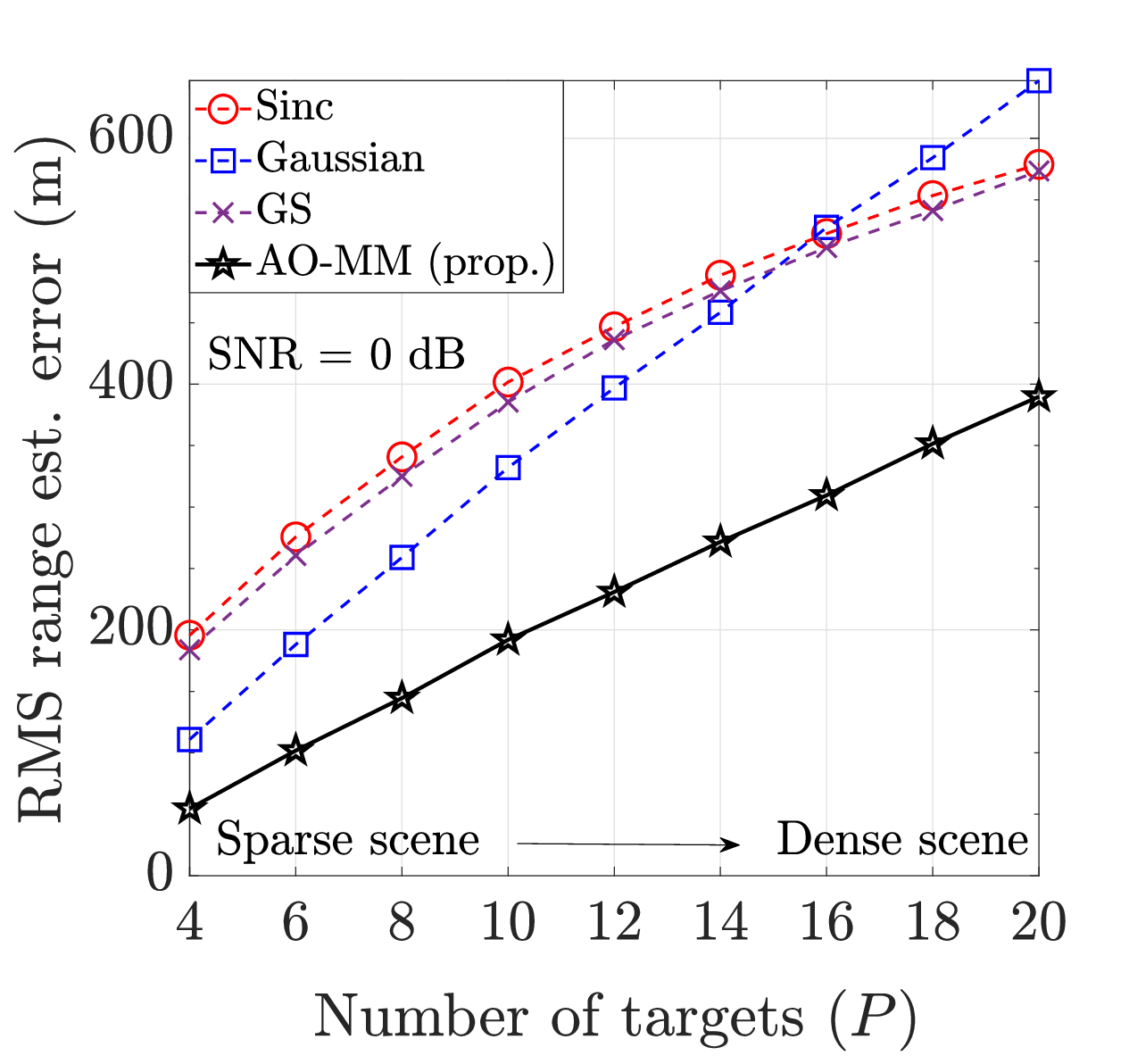}
    \label{Fig:6a}
    \vspace{-4mm}
    \caption{Range.}
\end{subfigure}
\begin{subfigure}[b]{0.49\linewidth}
    \includegraphics[width=\linewidth,height=3.6cm]{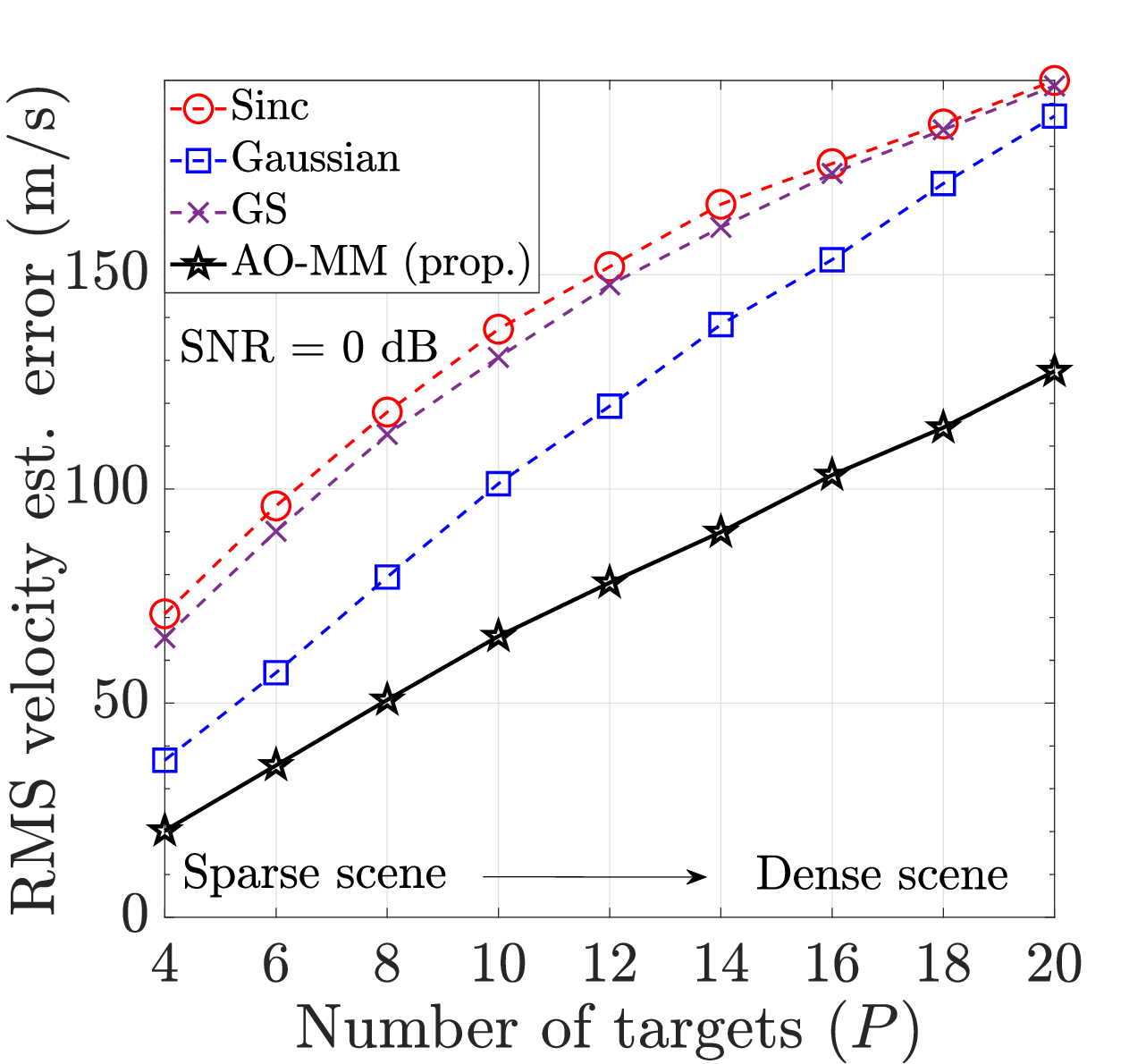}
    \label{Fig:6b}
    \vspace{-4mm}
    \caption{Velocity.}
\end{subfigure}
\caption{RMS range and velocity estimation errors versus number of targets $P$ at SNR $=0$\,dB.}
\label{Fig6}
\end{figure}

{\em Multi-target scenes:} We next consider multi-target~scenes with $P$ targets uniformly distributed over the scene region $\Omega=[0,5]\,\mu$s$\,\times\,[-1,1]$\,kHz. 
We use a detection window of $[-2,7]\,\mu$s$\,\times\,[-1.4,1.4]$\,kHz, which captures the spread of $A_{s,s}(\tau,\nu)$ for all filters. Fig. \ref{Fig5} shows the RMS range and velocity estimation errors as a function of SNR
for  $P=12$ targets. At low SNRs, the performance is noise-limited and all filters perform comparably. As the SNR increases, the sinc, GS and Gaussian filters quickly saturate at error floors set by inter-target interference, caused by sidelobe masking in the case of sinc and GS filters and by overlapping main lobes in the case of Gaussian filter. In contrast, the proposed AO-MM filter continues to improve up to an SNR of about $0$\,dB and saturates at a floor about $42\%$ (range) and $34\%$ (velocity) below those of sinc, Gaussian, and GS filters.

In Fig. \ref{Fig6}, for the same target region $\Omega$, we vary the number of targets from $P=4$, which corresponds to a sparsely populated scene, to $P=20$, which corresponds to a densely populated scene, at an SNR of $0$\,dB, where all filters are interference-limited. The estimation errors of all filters increase with $P$ owing to increasing inter-target interference. The sinc and GS filters are limited by sidelobe masking throughout, while the Gaussian filter, though better than sinc and GS filters for small $P$, degrades rapidly as the wide main lobes begin to overlap, eventually exceeding the sinc filter in range error. The proposed AO-MM filter achieves the lowest estimation error for both small (sparse) and large (dense) values of $P$, and its advantage widens as the scene becomes denser, making it superior across target densities.

\section{Conclusions}
\label{sec:conclusions}
Motivated by the superior communication performance achieved by optimizing the Zak-OTFS pulse shaping filter under the MM filter design framework in \cite{MM_zak}, here we optimized the MM prototype filter that achieved superior radar sensing performance. The approach we adopted is to optimize the coefficients of the MM prototype filter in such a way that the sidelobe levels in the ambiguity function are substantially lowered while keeping the main-lobe width narrow, without incurring time and bandwidth expansion. The main lobe of the resulting AO-MM filter is significantly narrower than that of the Gaussian filter, and its PSLR and ISLR are far below those of the sinc and GS filters, inheriting their best attributes. Simulation results showed that the proposed AO-MM filter achieved significantly better range and velocity estimation performance compared to those of the sinc, Gaussian, and GS filters across both sparsely and densely populated multi-target scenes. Future work could consider optimization of MM and other flexible filters for integrated sensing and communication. 

\appendix[Proof of Theorem 1]
\section{Proof of Theorem~\ref{thm:AF}}
\label{app:Axx}
Since $p_{\tau}(\cdot)$ and $p_{\nu}(\cdot)$ are real and even, the MM filter satisfies $w_{\mathrm{tx}}(\tau,\nu)=w_{\mathrm{tx}}^{*}(-\tau,-\nu)$, and its self-ambiguity takes the separable form in (\ref{eq:Axx}). For $\xi\in\{\tau,\nu\}$, define
\begin{equation}
\Psi_{\xi}(u,f)\triangleq
\int_{-\infty}^{\infty} w_{\xi}(x)\,w_{\xi}(u-x)\,e^{j2\pi f(u-x)}\,dx.
\label{eq:Psi_def}
\end{equation}
The delay integral in (\ref{eq:Axx}) is $\Psi_{\tau}(t_{n},\nu)$, while the Doppler integral, using $x=f_{m}-\nu'$, becomes $\Psi_{\nu}(f_{m},n\tau_{\mathrm{p}})$, giving (\ref{eq:AF_thm}).
To evaluate (\ref{eq:Psi_def}), using $\cos(x+\pi\ell)=(-1)^{\ell}\cos(x)$ and Euler's formula in (\ref{eq:mm_pulse}), along with $\sqrt{X_{\xi}}=\sqrt{\zeta_{\xi}\rho_{\xi}}$, we obtain
\begin{equation}
w_{\xi}(x)=\kappa_{\xi}\sqrt{\zeta_{\xi}\rho_{\xi}}\,
\operatorname{rect}(\rho_{\xi}x)
\sum_{k=-(g-1)}^{g-1}a_{k}\,e^{j2\pi k\rho_{\xi}x},
\label{eq:w_series}
\end{equation}
with $a_{k}$ as in (\ref{eq:a_coeff}). Substituting (\ref{eq:w_series}) into (\ref{eq:Psi_def}) and interchanging the finite sums and the integral,
\begin{eqnarray}
\hspace{-6mm}
\Psi_{\xi}(u,f)
& \hspace{-2mm} = & \hspace{-2mm} \kappa_{\xi}^{2}\zeta_{\xi}\rho_{\xi}
\sum\nolimits_{k,q=-(g-1)}^{g-1}a_{k}a_{q}\,e^{j2\pi(\rho_{\xi}q+f)u}
\nonumber \\
& \hspace{-10mm} & \hspace{-10mm} \times
\int\operatorname{rect}(\rho_{\xi}x)\operatorname{rect}\big(\rho_{\xi}(u-x)\big)\,
e^{j2\pi\beta_{kq}x}\,dx,
\label{eq:Psi_int}
\end{eqnarray}
where $\beta_{kq}\triangleq\rho_{\xi}(k-q)-f$. The two windows are supported on $[-1/(2\rho_{\xi}),1/(2\rho_{\xi})]$ and $[u-1/(2\rho_{\xi}),u+1/(2\rho_{\xi})]$, so their intersection $[L,U]$ has midpoint $(U+L)/2=u/2$ and length $U-L=\Delta_{\xi}(u)$ (empty for $|u|\ge 1/\rho_{\xi}$). Therefore,
\begin{equation}
\int_{L}^{U}e^{j2\pi\beta_{kq}x}\,dx
=\Delta_{\xi}(u)\,e^{j\pi\beta_{kq}u}\,
\operatorname{sinc}\!\big(\beta_{kq}\Delta_{\xi}(u)\big),
\label{eq:rect_int}
\end{equation}
which holds for all $\beta_{kq}$, including $\beta_{kq}=0$. Substituting (\ref{eq:rect_int}) into (\ref{eq:Psi_int}) and combining the phases as $2\pi(\rho_{\xi}q+f)u+\pi\beta_{kq}u=\pi(\rho_{\xi}(k+q)+f)u$ yields (\ref{eq:Psi}), which completes the proof.

\end{document}